\documentclass[epjST]{svjour}
\usepackage{graphicx,amsmath,amssymb}
\usepackage{dcolumn} 
\usepackage{bm} 
\usepackage{comment}

\begin{document}

\title{Energy flow and dissipation in heterogeneous ensembles of coupled phase oscillators}
 
\author{{Ra\'ul I. Sosa\inst{1}\fnmsep\thanks{\email{iansosa996@gmail.com}} \and Dami\'an H. Zanette\inst{1,2}\fnmsep\thanks{\email{damian.zanette@ib.edu.ar}}}}

\institute{Centro At\'omico Bariloche and Instituto Balseiro,   Comisión Nacional de Energ\'{\i}a Atómica and Universidad Nacional de Cuyo,   8400 San Carlos de Bariloche, R\'{\i}o Negro, Argentina \and Consejo Nacional de Investigaciones Cient\'{\i}ficas y T\'ecnicas (CONICET), Argentina }

\abstract{
We present analytical expressions and numerical results for the rates of energy exchange between oscillators and with the environment in a heterogeneous ensemble of globally coupled mechanical phase oscillators. The system is in stationary motion under the combined action of an external harmonic excitation, coupling, and friction. Individual moments of inertia and friction coefficients are different between oscillators. Three dynamical regimes, with different degrees of entrainment with the external excitation, are characterized. In two of these regimes, the rates of energy exchange show nontrivial dependence on the  moments of inertia and friction coefficients, suggesting that the transfer of power between different parts of the ensemble can be manipulated by a convenient choice of the individual parameters.
}  
\maketitle
 
\section{Introduction}
\label{intro}

Since the inception of complex systems as a focus of interest within Physics, ensembles of coupled oscillators have provided one of the main archetypes for the study of the emergence of collective behavior in systems formed by many interacting dynamical elements. Coherent motion induced by coupling --namely, mutual synchronization of various degrees between single oscillations-- is identified as a paramount example of the kind of phenomena brought about by the interplay of individual dynamics and interactions \cite{sync1,mmz}. The study of collective behavior in large ensembles of coupled oscillators has been originally motivated by the observation of coherent dynamics in biological systems, ranging from molecular complexes and cell tissues to populations of insects and other living beings \cite{Winf1}. The basic mechanism underlying  synchronization in oscillator ensembles, initially proposed by Winfree \cite{Winf0} and later formalized by Kuramoto \cite{Kura}, is directly inspired in the reciprocal action of ``biological clocks''.

On the other hand, although the first recognized historical observation of synchronization was reported for two pendulum clocks (in 1665, by Christiaan Huygens \cite{sync1}), ensembles of interacting mechanical oscillators have received relatively little attention concerning their feasibility of becoming synchronized. The joint motion of mechanical oscillators, however, not only provides an effective theoretical illustration of coherent evolution in dynamical systems, but is also an important matter of consideration in many technological applications. In particular, the control of energy exchange between different components of a mechanical device, and between the device and its surroundings, is of high relevance to the functioning of a broad class of machines and instruments. Examples that involve coupled oscillators are, for instance, equipment for energy harvesting \cite{9,2}, pacemakers, and sensors \cite{10,12,11}.   

In a recent contribution \cite{nos}, we have studied energy exchange in an ensemble of globally coupled mechanical oscillators, in a situation where the system is excited by an external harmonic force applied to one of the oscillators. Friction forces, whose intensity varied from oscillator to oscillator, made it possible to reach a state of stationary motion where we characterized the flow of energy between different oscillators and with the environment. The conservative part of the dynamics, which encompassed coupling between oscillators, was described by the mean-field Hamiltonian \cite{ruffo,daux}, where the canonical coordinates of each oscillator are its phase and the conjugate (angular) momentum. 

Here, we generalize the analysis to the case where both the friction coefficients and the moments of inertia are different between oscillators. We show that this heterogeneity induces a varied dependence of the rates of energy exchange on the individual parameters of each element, which can even control the direction in which energy flows in different parts of the system. Energy exchange is characterized in three disparate regimes of coherent motion, depending on the degree of entrainment with the external excitation: full and partial synchronization, and desynchronization. In Sect.~\ref{sec2},  after introducing our model, we describe the main dynamical properties of the three regimes. In Sect.~\ref{sec3}, approximate analytical solutions for the equations of motion in each regime are presented. Using these solutions, in Sect.~\ref{sec4} we calculate the respective rates of energy exchange, and present a comparison with results obtained from direct numerical integration of the equations. Finally, conclusions are outlined in Sect.~\ref{sec5}.

\section{Globally coupled non-identical phase oscillators} \label{sec2}

Our system consists of an ensemble of $N$ phase oscillators subjected to global coupling, friction, and an external harmonic excitation applied to one of the oscillators --which, without generality loss, we identify as oscillator $1$. The dynamical state of  each oscillator is fully characterized by its phase $\theta_n (t) \in [0,2\pi)$ and the corresponding conjugate momentum $p_n(t)$. Coupling is given by a pair interaction which derives from the Hamiltonian mean-field (HMF) model,
\begin{equation} \label{ham}
{\cal H} = \sum_{n=1}^N \frac{p_n^2}{2I_n} -\frac{K}{N}\sum_{n<m}^N \cos (\theta_m-\theta_n) ,
\end{equation}
where $I_n$ is the moment of inertia of oscillator $n$, and $K$ is the coupling strength. We recall that, owing to its minimalist nature, the HFM model has been widely applied  to the study of collective behavior in Hamiltonian systems \cite{ruffo,daux}. Here, we use it as a convenient stylized representation of any ensemble of coupled mechanical (nonlinear) oscillators. 

To the equations of motion derived from the Hamiltonian in Eq.~(\ref{ham}), we add friction forces proportional to the velocities $\dot \theta_n$, and the external harmonic excitation, getting 
\begin{equation} \label{Newton}
I_n \ddot \theta_n = \frac{K}{N} \sum_{m=1}^N \sin (\theta_m -\theta_n) -  \gamma_n \dot \theta_n + F \sin (\omega t-\theta_1) \delta_{n1},
\end{equation}
for $n=1,\dots, N$, where $\gamma_n$ are the friction coefficients. In the last term, $F$ and $\omega$ are the amplitude and frequency of the external excitation, and $\delta_{n1}$ is Kronecker's delta. Note that  this forcing is equivalent to the interaction of oscillator $1$ with an oscillator of prescribed phase $\omega t$, having the same dependence on phases as the coupling described by the first term in the right-hand side.

Equations (\ref{Newton}) adopt a more tractable form if we introduce a Kuramoto-like order parameter \cite{Kura}
\begin{equation} \label{Z}
 R \, {\rm e}^{i \Theta} = \frac{1}{N-1} \sum_{n=2}^N {\rm e}^{i \theta_n}.
\end{equation}
Because of the special dynamical status of oscillator $1$, which experiences the action of the external force, we have restricted the sum in Eq.~(\ref{Z}) to $n>1$. Hereafter, the group of oscillators with $n=2,\dots, N$ is called the $\Omega$ {\em set}. 

It is convenient, moreover, to rescale time in such a way that the frequency of the excitation equals unity. This amounts to renaming $\omega t\to t$, $\omega^{-2} K \to K$, $\omega^{-1} \gamma_n \to \gamma_n$, and $\omega^{-2} F \to F$  for all $n$. With these definitions, Eqs.~(\ref{Newton}) finally read
\begin{equation} \label{mot1}
     I_1 \ddot \theta_1 = \bar K R \sin (\Theta -\theta_1)-  \gamma_1  \dot \theta_1 +  F \sin ( t-\theta_1)
\end{equation}
for oscillator 1, and
\begin{equation} \label{motn}
 I_n\ddot \theta_n = \bar K R \sin (\Theta -\theta_n)-   \gamma_n\dot \theta_n+  k \sin (\theta_1-\theta_n)
\end{equation}
for $n=2,\dots ,N$, with $\bar K =(N-1) K/N$, and $k=K/N$. 

The instantaneous power exchanged by oscillator $n$ with the rest of the system is given by the product $I_n \dot \theta_n\ddot \theta_n$. As the oscillator receives or releases energy, this quantity is respectively positive or negative. Its average over time, $w_n  =I_n  \langle \dot \theta_n\ddot \theta_n \rangle_t$, can be divided into contributions coming from the three mechanisms that drive the dynamics, namely, coupling, friction, and the external excitation. Specifically, we write $w_n = w_n^\Omega+w_n^\Gamma +w_n^F$, where
\begin{equation} \label{6}
    w_n^\Omega =  \bar K \langle R \sin (\Theta -\theta_n)\dot  \theta_n \rangle_t
\end{equation}
is the rate of energy exchange between oscillator $n$ and the $\Omega$ set, and
\begin{equation}
    w_n^\Gamma = -  \gamma_n \langle \dot \theta_n^2 \rangle_t,
\end{equation}
equals the power lost by friction. Moreover, 
\begin{equation}
w_1^F = F \langle  \sin (t-\theta_1) \dot \theta_1 \rangle_t 
\end{equation}
corresponds to the power received or released by oscillator $1$ from or towards the source of  external forcing, and
\begin{equation} \label{9}
w_n^F = k \langle  \sin (\theta_1-\theta_n)  \dot \theta_n \rangle_t,
\end{equation}
for $n=2,\dots N$, is the rate of energy exchange of an oscillator in the $\Omega$ set with oscillator $1$. 

\begin{figure}[ht]
\includegraphics[width= \columnwidth]{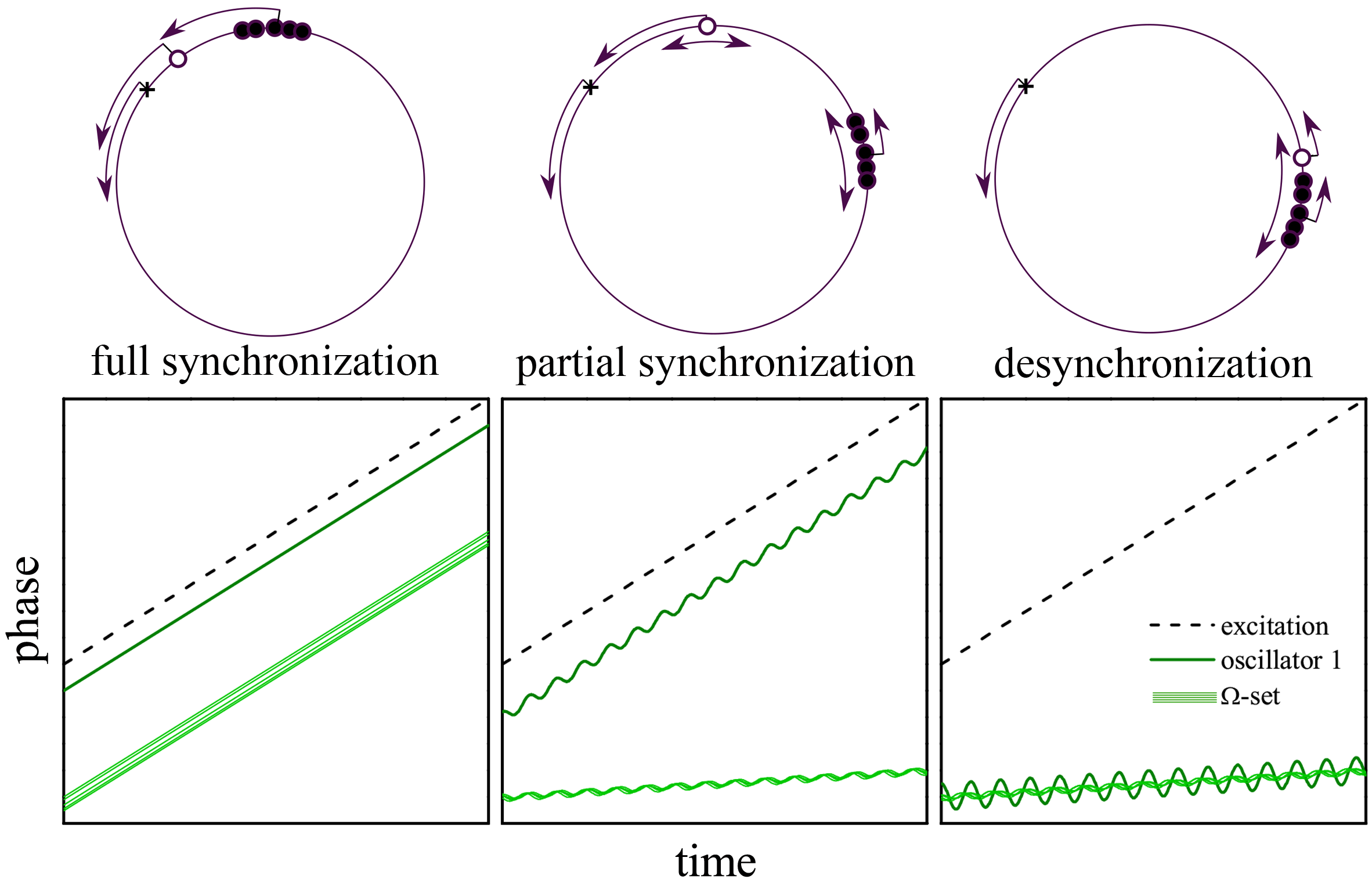} 
\caption{\label{fig1} Upper row: Illustrative diagram of the position and the velocity of oscillators around the circle of phases, in the three regimes of collective behavior described in the text. The cross indicates the phase of the external excitation. Open and full dots represent oscillator $1$ and the $\Omega$ set, respectively. Single arrows stand for the frequency of excitation and the average frequency of oscillators, and double arrows denote small oscillations around the average motion. Lower row: Schematic representation of the time dependence of phases for the excitation, oscillator $1$, and the $\Omega$ set, as indicated in the legend.}
\end{figure}

As discussed in our previous publication \cite{nos}, the present system is expected to exhibit three qualitatively different stationary regimes of collective behavior, depending on the parameters which control the dynamics. These regimes are characterized by diverse degrees of synchronization with the external excitation. It is important to remark that they are not necessarily separated by sharp boundaries in parameter space, but rather belong to well-differentiated ranges of $K$, $F$, and of the individual parameters $I_n$ and $\gamma_n$. The three regimes are schematized in Fig.~\ref{fig1}. 

When the external excitation is sufficiently strong as compared to the action of coupling and friction, oscillators reach a state of {\it full synchronization}. In this regime,  the whole ensemble moves rigidly with the same frequency as the forcing. Different oscillators, however, have different phases, depending on their individual moments of inertia and friction coefficients. 

If the external excitation remains dominant but friction forces overcome the effects of coupling, the average motion of oscillator $1$ is still synchronized with the forcing. In contrast, the $\Omega$ set detaches itself from synchronized motion and, on the average, moves at a much slower pace. Moreover, in this regime of {\em partial synchronization}, all oscillators perform small-amplitude oscillations around their average phase drift. 

Finally, when the external force is weak as compared to coupling and friction, the excitation is unable to entrain the ensemble and {\em desynchronization} follows. In this situation, oscillator $1$ becomes aggregated to the $\Omega$ set, and the whole ensemble moves with a frequency much smaller than that of the forcing, also modulated by small oscillations. 

In the following, we provide a quantitative description of these three stationary dynamical regimes, within some simplifying assumptions that make the problem analytically tractable. Results allow us to explicitly calculate the rates of energy exchange  $w_n^\Omega$, $w_n^\Gamma$, and $w_n^F$ of Eqs.~(\ref{6}) to (\ref{9}). These quantities are compared with estimations obtained from numerical solution of the equation of motion, illustrating the variety of behaviors that energy exchange can adopt all over the system. 

\section{Analytical treatment of collective dynamics} \label{sec3}

The equations of motion (\ref{mot1}) and (\ref{motn}) can be dealt with analytically if the standard deviation of the friction coefficients and the moments of inertia over the ensemble, $\sigma_\gamma$ and $\sigma_I$ respectively, are sufficiently small as compared with the coefficients themselves. Numerical integration of the equations show that, under these conditions, the dispersion of phases between the oscillators in the $\Omega$ set is small. Based on this evidence, we assume  that the individual phases $\theta_n$  differ from their arithmetic average $\langle \theta \rangle=(N-1)^{-1} \sum_{n=2}^N \theta_n$ by   quantities $\epsilon_n =\theta_n - \langle \theta \rangle$ which are proportional to $\sigma_\gamma$ and $\sigma_I$. Neglecting terms of order $\sigma_\gamma^2$ and $\sigma_I^2$ in  Eq.~(\ref{Z}), we get $R\, \exp (i \Theta) = \exp (i \langle \theta \rangle)$ or, equivalently, 
\begin{equation}
R=1, \ \ \ \ \ \Theta=\langle \theta \rangle.
\end{equation}
Moreover, if we disregard statistical correlations between friction coefficients and velocities on one side, and moments of inertia and accelerations on the other, we can write
\begin{equation}
\frac{1}{N-1} \sum_{n=2}^N \gamma_n \dot \theta_n = \langle \gamma \rangle \langle \dot \theta \rangle , \ \ \ \ \ \frac{1}{N-1} \sum_{n=2}^N I_n \ddot \theta_n = \langle I \rangle \langle \ddot \theta \rangle ,
\end{equation}
where $\langle \cdot \rangle$ stands for averages over the $\Omega$ set. Naturally, the validity of these assumptions is to be assessed from comparison with numerical results, which we postpone to next section. 

Replacing our simplifying hypotheses in Eqs.~(\ref{mot1}) and (\ref{motn}), we get a pair of coupled equations of motion for the phase of oscillator $1$ and for the average phase of the $\Omega$ set, $\theta_1$ and $\Theta$, of the form
\begin{eqnarray}
   I_1 \ddot \theta_1  &=& \bar K  \sin (\Theta -\theta_1)-\gamma_1\dot \theta_1 +  F \sin (t-\theta_1),\label{m1a}  \\ \nonumber  \\
   \langle I \rangle \ddot \Theta   &=& -\langle \gamma \rangle \dot \Theta + k \sin (\theta_1 - \Theta).\label{mna}
\end{eqnarray}
The equation of motion for the individual phase differences in the $\Omega$ set, $\epsilon_n=\theta_n-\Theta$, is obtained  by dividing Eq.~(\ref{motn}) by $I_n$, subtracting Eq.~(\ref{mna}) divided by $\langle I \rangle$, and approximating to the first order in $\epsilon_n$:
\begin{equation} \label{mea}
\ddot \epsilon_n = -\bar K_n  \epsilon_n+ \frac{ K_n-K}{N} \sin(\theta_1 - \Theta) -(\gamma_n -\langle \gamma \rangle) \dot \Theta -\gamma_n   \dot \epsilon_n.
\end{equation}
Here, for the sake of brevity in notation, we have renamed $\gamma_n/I_n\to \gamma_n$, $\langle \gamma \rangle / \langle I\rangle \to\langle \gamma \rangle$, and $K/ \langle I\rangle\to K$. Also, we have defined $K_n=K/I_n$, $\bar K_n=\bar K/I_n$. Note that Eq.~(\ref{mea}) couples the dynamics of $\epsilon_n$ with those of $\theta_1$ and $\Theta$.

Equations (\ref{m1a}) to (\ref{mea}) constitute our approximation to deal analytically with the equations of motion (\ref{mot1}) and (\ref{motn}). In the following, we study their solution in the three synchronization regimes pointed out in Sect.~\ref{sec2}.

\subsection{Full synchronization}

For the whole ensemble to synchronize with the external excitation it is necessary, first, that oscillator $1$ becomes entrained by the harmonic force. This requires that, in Eq.~(\ref{mot1}), the last term in the right-hand side dominates over the other two, $F\gg \bar K, \gamma_1$. Under these conditions, for long times, the phases of oscillator $1$ and the external excitation differ by a small quantity
\begin{equation} \label{theta1}
\eta (t)=t -\theta_1 (t) .
\end{equation}
To the first significant order in $ \eta$, Eq.~(\ref{mot1}) becomes
\begin{equation} \label{m1r1}
\ddot  \eta  = \bar K  \sin (\tau-\Theta)-\gamma_1 ( \dot \eta  - 1) -  F \eta,
\end{equation}
while from Eq.~(\ref{motn}) we get
\begin{equation} \label{mnr1}
   \ddot \Theta = -\langle \gamma \rangle \dot \Theta+ k \sin (t- \Theta).
\end{equation}
This is the equation of motion for a mechanical oscillator of phase $\Theta$ subjected to friction and an external harmonic force of unitary frequency. It can be readily shown \cite{mmz,nos} that, for long times, its solution becomes synchronized to the force if $k > \langle \gamma \rangle$. Thus, this is the second condition that defines the regime of full synchronization in our oscillator ensemble.  The asymptotic stationary solution for the average phase in the $\Omega$ set is 
\begin{equation} \label{Theta}
    \Theta= t- \arcsin \frac{\langle \gamma \rangle}{k} .
\end{equation}
Replacing this result in Eq.~(\ref{m1r1}), and taking into account the first condition that defines the regime of full synchronization, we consistently verify that $|\eta| \ll 1$ for long times. 

Finally, substituting Eq.~(\ref{Theta}) in Eq.~(\ref{mea}), we obtain the individual deviations of oscillators in the $\Omega$ set with respect to their average phase: 
\begin{equation} \label{epsr1a}
    \epsilon_n = \frac{\langle \gamma \rangle}{\bar K_n} \left(\delta K_n -  \delta \gamma_n \right),
\end{equation}
with
\begin{equation}
\delta K_n = \frac{K_n-K}{K}, \ \ \ \ \ \delta \gamma_n = \frac{\gamma_n-\langle \gamma \rangle}{\langle \gamma \rangle}.
\end{equation}
Note that $\epsilon_n$ results from the combination of the deviations of individual coupling constants $K_n$ and friction coefficients $\gamma_n$ with respect to their average values. Depending on their relative sign, these deviations can mutually reinforce their effect or compensate each other.

\subsection{Partial synchronization}

When $F\gg \bar K, \gamma_1$ and $\langle \gamma \rangle  > k$, Eq. (\ref{theta1}) and Eq. (\ref{m1r1}) still hold and, therefore, oscillator $1$ is entrained by the external excitation and moves with the same frequency, but the $\Omega$ set does not synchronize with oscillator $1$. This situation defines the regime of partial synchronization, for which we are able to provide an analytical description in the limit $\langle \gamma \rangle \gg k$. On the basis of numerical evidence, as advanced in Sect.~\ref{sec2}, we propose a stationary solution where the average phase in the $\Omega$ set moves with a small constant drift velocity $\nu$ and, at the same time, oscillates around this uniform motion with a small amplitude and a frequency induced by its interaction with oscillator $1$, namely,
\begin{equation} \label{anz1}
    \Theta (t) = A \cos [(1-\nu) t - \Psi]+\nu t.
\end{equation}
Substitution into  Eq.~(\ref{mnr1}) gives, to the leading order in $\nu$ and $A$,
\begin{equation} \label{A1b}
A = \frac{k}{\sqrt{1+\langle \gamma \rangle^2}}
\end{equation}
where we have neglected higher-harmonic contributions coming from the last term in the right-hand side. Furthermore, averaging the equation over the fast oscillation of frequency $1-\nu$, we find
\begin{equation} \label{nu1b}
\nu = \frac{1}{2}A^2 = \frac{k^2}{2(1+\langle \gamma \rangle^2)}
\end{equation}
for the drift velocity. Consistently with our approximation to the regime of partial synchronization, we have $A,\nu\ll 1$. Moreover,   replacing the stationary solution for the average phase in the $\Omega$ set into Eq. (\ref{m1r1}), and using the defining conditions of this regime, we find that the approximation $|\eta| \ll 1$ holds. This implies that oscillator $1$ is entrained by the external excitation, which makes our analytical description consistent.

For asymptotically long times, the individual deviations from the average motion of the $\Omega$ set are now time-dependent:
\begin{equation} \label{eps1b}
    \epsilon_n (t) = A \sqrt{\frac{(\delta K_n)^2+\langle \gamma \rangle^2 (\delta K_n  -  \delta \gamma_n)^2}{\left(\bar K_n-1 \right)^2+\gamma_n^2}} \cos(t-\psi_n),
\end{equation}
with $A$ given by Eq.~(\ref{A1b}). The phase shifts $\Psi$ in Eq.~(\ref{anz1}) and $\psi_n$ in Eq.~(\ref{eps1b}) can also be given explicit expressions. However, since these algebraically cumbersome formulas are not involved in the rates of energy exchange we are interested in, we avoid reporting them for brevity.    

Note that, in contrast with the result of Eq.~(\ref{epsr1a}) in the regime of full synchronization, the deviations $\epsilon_n$ in Eq.~(\ref{eps1b}) cannot be suppressed by mutually compensating $\delta K_n$ and $\delta \gamma_n$. In the regime of partial synchronization, a dispersion in the values of $K_n$ --i.e. in the moments of inertia-- is enough to induce non-trivial dynamics inside the $\Omega$ set.

\subsection{Desynchronization} \label{ssec3C}

When the condition on the external excitation that defines the regimes of full and partial synchronization is inverted, $F\ll \bar K , \gamma_1$, the oscillator ensemble cannot be entrained by the force, and moves out of synchrony. In this regime, at long times, both oscillator $1$ and the $\Omega$ set exhibit a slow drift with the same velocity, and small oscillations around this motion. Our proposal for the solutions to Eqs.~(\ref{mot1}) and (\ref{motn}) is now
\begin{equation} \label{anzr2}
\begin{array}{rl}
\theta_1(t) &= a_1 \cos[(1-\nu) t-\psi_1]+\nu t, \\   \\
\Theta(t) &= A \cos[(1-\nu) t-\Psi]+\nu t+\Theta_0. 
\end{array}
\end{equation}
To the leading order in amplitudes and velocities, we get
\begin{eqnarray}
 a_1 &=& \frac{F}{\sqrt{\left[\bar q\bar K-1+ q ( N-1)\right]^2+\left(\bar q \gamma_1 +q \Gamma \right)^2}} , \label{a1}\\
  A &=& \frac{\sqrt{q}F}{\sqrt{\left[\bar q\bar K-1+ q ( N-1)\right]^2+\left(\bar q \gamma_1 +q \Gamma \right)^2}}, \label{A2}
\end{eqnarray}
and
\begin{equation} \label{nu2}
    \nu= \frac{1}{2} A^2 + \frac{\bar q \gamma_1}{2 \Gamma} a_1^2,
\end{equation}
with $\Gamma=\sum_{n=2}^N \gamma_n$ and
\begin{equation} \label{r}
q\equiv \frac{k^2}{(k-1)^2 + \langle \gamma \rangle^2},  \ \ \ \bar q=1-q.
\end{equation}
The individual deviations $\epsilon_n$ have  the same form as in Eq.~(\ref{eps1b}), with $A$ now given by Eq.~(\ref{A2}). As in the regime of partial synchronization, the phase shifts in Eqs.~(\ref{anzr2}) and in the expression for $\epsilon_n$ can be given explicit expressions, which we do not report here.

\section{Rates of energy exchange: analytical and numerical results} \label{sec4}

Having obtained approximate stationary solutions for the equations of motion in the three regimes of collective dynamics of our system, we can now proceed to calculate the rates of energy flow and dissipation, $w_n^\Omega$, $w_n^\Gamma$, and $w_n^F$, given by Eqs.~(\ref{6}) to (\ref{9}). The procedure amounts to substituting the solutions found in the preceding section and, within the same approximation order, compute the time averages which define the quantities $w_n^\Omega$, $w_n^\Gamma$,  and $w_n^F$. As we show in the following, the results can be given explicit analytical expressions.

At the same time,  we compare our analytical results with estimations of the same quantities obtained from direct numerical integration of the equations of motion (\ref{Newton}). This integration was performed using a standard Runge–Kutta fourth-order ODE solver implemented in C++. At every time step, the sum of forces acting over each oscillator was calculated in GPU by means of the Thrust parallel algorithms library \cite{Thrust}. With the aim of maintaining consistency and accuracy in the computation, we strictly adhered to the following protocol. (i) Set the total integration time to ten times the maximum  ratio between   moment of inertia and friction coefficient over the ensemble,  $10 \max_n \{I_n/\gamma_n\}$. This setting satisfactorily ensures that transients have elapsed and that stationary oscillations have been reached. (ii) Choose a sufficiently small time discretization, so that the total deviation in the conservation of energy due to the numerical errors remains below $0.1$\%. (iii)  Store in video memory the data for the position and the velocity of each oscillator during ten successive oscillations in the stationary state. (iv) Calculate the time average of the different rates of energy exchange using the stored data, by means of GPU parallel computation. In all cases studied below, the relative discrepancy between analytical and numerical results consistently remained within a few percent.

From the analytical solutions in the regime of full synchronization, we get
\begin{eqnarray}
    w_1^\Omega &=&  -(N-1) \langle \gamma \rangle, \nonumber  \\
    w_1^\Gamma &=& -\gamma_1,  \label{w1a1} \\ 
    w_1^F &=&  (N-1)  \langle \gamma \rangle+\gamma_1 , \nonumber
\end{eqnarray}
for oscillator 1, and 
\begin{eqnarray}
    w_n^\Omega &=& 
    \frac{K}{K_n}  \langle \gamma \rangle (\delta \gamma_n -\delta K_n),  \nonumber \\
    w_n^\Gamma &=& - \frac{K}{K_n}\gamma_n, \label{w1a} \\ 
    w_n^F &=& \langle \gamma \rangle, \nonumber
\end{eqnarray}
for $n=2,\dots,N$.  It can be shown that $ w_n^\Omega + w_n^\Gamma + w_n^F =0$ for all $n$, as expected to happen in stationary motion, where the incoming and the outgoing mechanical power must compensate each other for each oscillator. 

\begin{figure}[ht]
\includegraphics[width= 0.8\columnwidth]{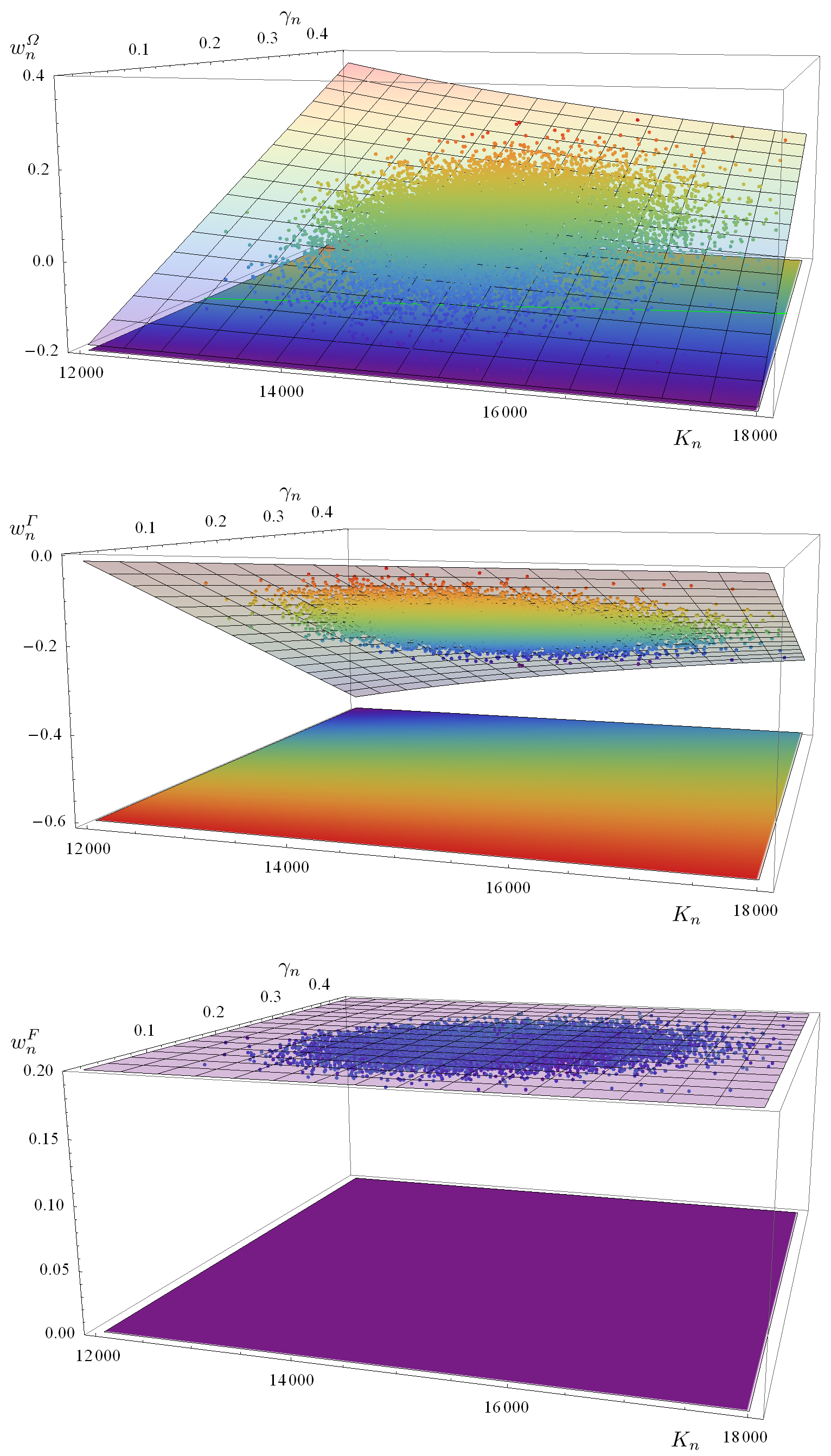} 
\caption{\label{fig2} Rates of energy exchange, $w_n^\Omega$, $w_n^\Gamma$, and $w_n^F$, in the regime of full synchronization as functions of the friction coefficients $\gamma_n$ and the coupling constants $K_n$ of individual oscillators in the $\Omega$ set. The gridded surfaces correspond to the analytical results of  Eqs.~(\ref{w1a}), and the  dots show numerical results for $N=2 \times 10^4$, $\bar K=1.5 \times 10^4$ and $F=10^5$. The friction coefficients and moments of inertia were drawn from a Gaussian distribution with mean values $\langle \gamma \rangle =0.2$ and $\langle I \rangle =1$, and standard deviations $\sigma_\gamma =0.05$, $\sigma_I =0.1$ respectively. The bright line on the horizontal surface of the uppermost panel indicates the curve in the ($K_n,\gamma_n$)-plane where $w_n^\Omega=0$.}  
\end{figure}

In this regime, the rates of energy exchange depend linearly on the friction coefficients and are inversely proportional to the constants $K_n$ --i.e. they are linear on the moments of inertia $I_n$ (see Sect.~\ref{sec3}). Moreover, the exchange of energy between oscillators in the $\Omega$ set, measured by $w_n^\Omega$, can change sign depending on the deviations of $\gamma_n$ and $K_n$ with respect to their averages. These deviations, in turn, can reinforce or compensate each other; cf.~Eq.~(\ref{epsr1a}). Figure \ref{fig2} shows three-dimensional plots of $w_n^\Omega$, $w_n^\Gamma$, and $w_n^F$ for the $\Omega$ set over the plane spanned by $K_n$ and $\gamma_n$. Gridded surfaces correspond to the analytical results of Eqs.~(\ref{w1a}) and dots are numerical estimations for a system of $2\times 10^4$ oscillators, with the parameters indicated in the caption.

In the regime of partial synchronization,  we have
\begin{eqnarray}
    w_1^\Omega &=& -\frac{A^2}{2} (N-1) \langle \gamma \rangle ,\nonumber  \\
    w_1^\Gamma &=& -\gamma_1 ,  \label{w1b1}  \\ 
    w_1^F &=& \frac{A^2}{2} (N-1) \langle \gamma \rangle+ \gamma_1, \nonumber
\end{eqnarray}
for oscillator $1$,  where the amplitude $A$ is given by Eq.~(\ref{A1b}). Meanwhile, the rates of energy exchange for the $\Omega$ set are
\begin{eqnarray}
    w_n^\Omega &=& 
    \frac{A^2}{2} \frac{\bar K \langle \gamma \rangle}{(\bar K_n-1)^2+\gamma_n^2} \left[ \delta \gamma_n (\bar K_n -1) -\delta K_n (\bar K_n-1+\gamma_n /\langle \gamma\rangle)\right]    ,\nonumber  \\
    w_n^\Gamma &=& 
- \frac{A^2 }{2}  \frac{\gamma_n K/ K_n}{(\bar K_n-1)^2+\gamma_n^2}
[ (K_n-1)^2 - \langle \gamma\rangle ^2 \delta \gamma_n   \nonumber \\
& &+2 \delta K_n (\langle \gamma\rangle \gamma_n -\bar K_n+1) + (\delta K_n)^2 (1+\langle \gamma \rangle^2) ] , \label{w1b} \\
    w_n^F &=&  
\frac{A^2 }{2}  \frac{\langle \gamma \rangle}{(\bar K_n-1)^2+\gamma_n^2} [(\bar K_n-1)^2+\gamma_n^2-\delta \gamma_n (\bar K-1+\langle \gamma \rangle \gamma_n) \nonumber \\ 
& &+\delta K_n \gamma_n (1+\langle \gamma \rangle^2)/\langle \gamma \rangle] . \nonumber
\end{eqnarray}
Now, as a consequence of the approximations involved in our analytical approach to partial synchronization, $ w_n^\Omega + w_n^\Gamma + w_n^F =0$ to the first order in $\delta K_n$ only. 

\begin{figure}[ht]
\includegraphics[width= 0.8\columnwidth]{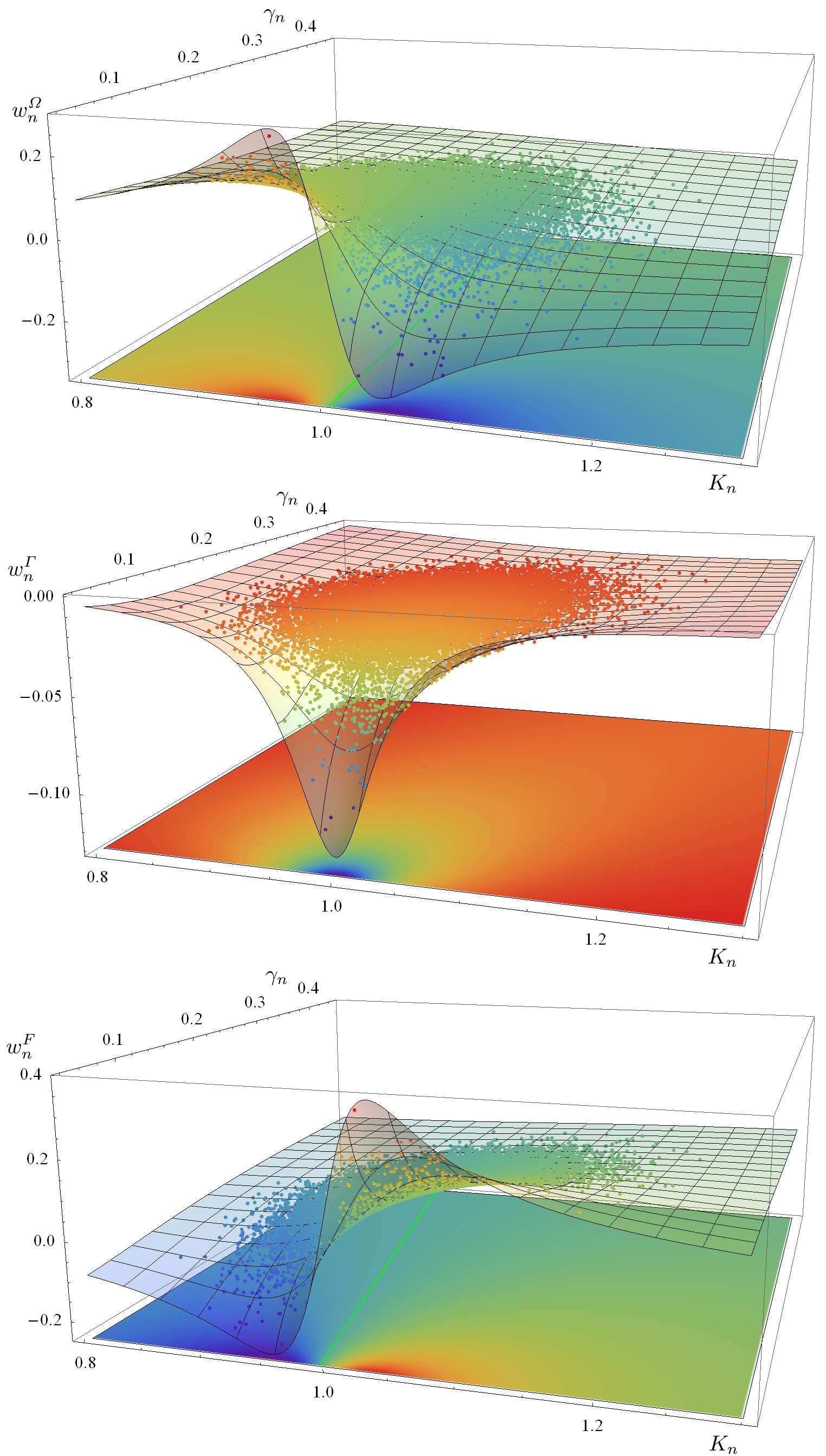} 
\caption{\label{fig3} As in Fig.~\ref{fig2}, for the regime of partial synchronization. Here, $\bar K=1$ and $F=10^3$. For clarity in the representation, scales in the vertical axes are multiplied by a factor $10^8$. Bright lines on the horizontal planes of the upper- and lowermost panels indicate the curves in the ($K_n,\gamma_n$)-plane where $w_n^\Omega=0$ and $w_n^F=0$, respectively.}
\end{figure}

Figure \ref{fig3} shows the quantities $w_n^\Omega$, $w_n^\Gamma$ and $w_n^F$ for the $\Omega$ set in the regime of partial synchronization. The first substantial difference with the case of full synchronization is that the rates of energy exchange are now much smaller (the vertical axes in the plots are amplified by a factor of $10^8$). This is due to the fact that, as implied by Eq.~(\ref{A1b}), oscillations have an amplitude of order $K/N$. As a consequence, for a given value of $K$, energy flow and dissipation are of order $N^{-2}$. 

However, the main feature in the rates of energy exchange for partial synchronization  resides in their non-trivial dependence on both the friction coefficient $\gamma_n$ and the coupling constant $K_n$. The profile of the rate of energy dissipation (middle panel of Fig.~\ref{fig3}) points to the existence of a resonance effect for $K_n \approx 1$, which becomes enhanced as $\gamma_n$ approaches zero.  Mathematically, the sharp downward peak of $w_n^\Gamma$ is a direct consequence of the denominator $(\bar K_n-1)^2+\gamma_n^2$ in Eq.~(\ref{w1b}), which is already present in the amplitude of individual departures from the $\Omega$ set, Eq.~(\ref{eps1b}). This resonance can be traced back to the equation of motion (\ref{mea}) for the deviation $\epsilon_n$ where, as demonstrated by the first term in the right-hand side, $\bar K_n$ plays the role of the squared natural frequency of $\epsilon_n$. From the third term, meanwhile, we see that $\epsilon_n$ is being forced harmonically by the $\Omega$ set, through its phase $\Theta$. This phase moves with unitary frequency as a collective response to the coupling with oscillator $1$, which is in turn driven by the external excitation. In short, $\epsilon_n$ behaves as a damped harmonic oscillator of squared frequency $\bar K_n$ subjected to an excitation of unitary frequency. Thus, it exhibits a resonance at  $\bar K_n=1$. Because of their more complicated dependence on $\bar K_n$, the rates $w_n^\Omega$ and $w_n^F$ exhibit a change of sign for $K_n \approx 1$, with conspicuous extrema at each side, at least, for small $\gamma_n$. This behavior resembles the dependence of the phase of oscillation near the resonance of a periodically forced, damped harmonic oscillator.

Finally, for the regime of desynchronization we get
\begin{eqnarray}
    w_1^\Omega &=& - (N-1) \langle \gamma \rangle \left( \frac{A^2}{2}  + \nu^2 \right),  \nonumber  \\
    w_1^\Gamma &=& -\gamma_1 \left( \frac{a_1^2 }{2}  + \nu^2 \right),  \label{w21} \\ 
    w_1^F &=& \frac{A^2}{2}  (N-1) \langle \gamma \rangle+\frac{a_1^2 }{2} \gamma_1 +\Gamma \nu^2 , \nonumber
\end{eqnarray}
for oscillator 1, with $a_1$, $A$, and $\nu$ given by Eqs.~(\ref{a1}) to (\ref{nu2}), respectively. For the $\Omega$ set, in turn,  we find
\begin{eqnarray}
        w_n^\Omega &=& -\bar K_n\frac{A^2}{2}\frac{\langle \gamma \rangle \delta r_n(\bar K_n-1)+\gamma_n \delta K_n}{(\bar K_n-1)^2+\gamma_n^2}-\langle \gamma \rangle \frac{K}{K_n}\delta r_n \nu^2,\nonumber \\ 
    w_n^\Gamma &=& - \frac{A^2 }{2}  \frac{\gamma_n K/ K_n}{(\bar K_n-1)^2+\gamma_n^2}[(\bar K_n-1)^2+\gamma_n^2+(\langle \gamma \rangle \delta r_n)^2+( \delta K_n)^2 \nonumber \\
    & & +\gamma_n \langle \gamma \rangle \delta r_n - (\bar K_n-1) \delta K_n ]  -\gamma_n \frac{K}{K_n} \nu^2,  \label{w2} \\ 
    w_n^F &=&  \frac{A^2}{2} \langle \gamma \rangle \left[ 1 +   \frac{\gamma_n \langle \gamma \rangle  \Delta r_n-\delta \gamma_n (\bar K_n-1)+\gamma_n \delta K_n/\langle \gamma \rangle}{(\bar K_n-1)^2+\gamma_n^2} \right] \nonumber \\
& &+\frac{K}{K_n}\nu^2 (\langle \gamma \rangle \delta r_n + \gamma_n)\nonumber
\end{eqnarray}
with $\delta r_n \equiv \delta K_n  - \delta \gamma_n$. Now, $ w_n^\Omega + w_n^\Gamma + w_n^F =0$ to the first order in $\delta K_n$ and $\delta \gamma_n$.

\begin{figure}[ht]
\includegraphics[width= .8\columnwidth]{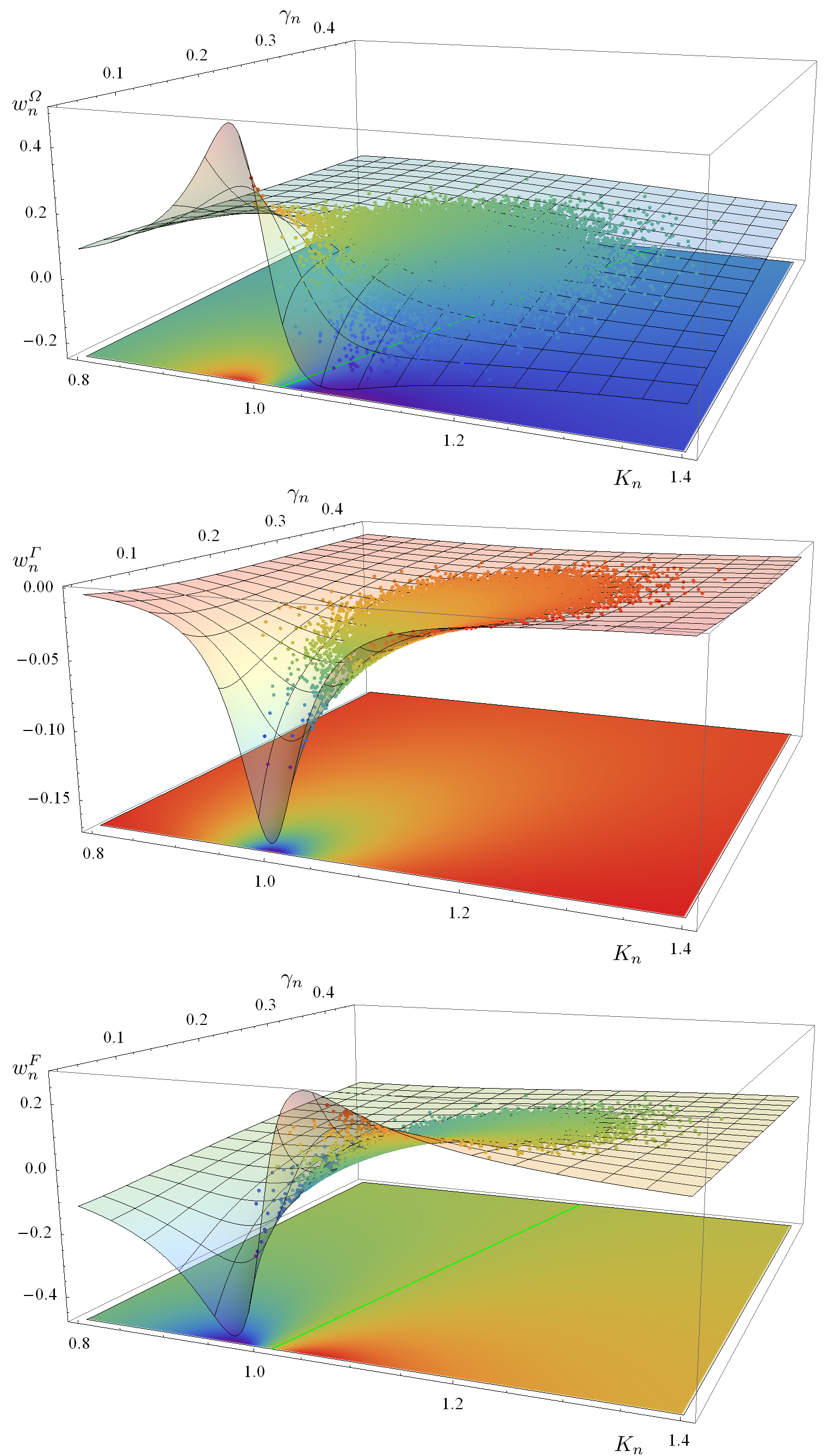} 
\caption{\label{fig4}As in Fig.~\ref{fig3}, for the regime of desynchronization. Here, $\bar K=1.1$ and $F=10^{-2}$. For clarity in the representation, scales in the vertical axes are multiplied by a factor $10^{12}$.}
\end{figure}

Figure \ref{fig4} shows our analytical and numerical results for the regime of desynchronization in the $\Omega$ set. With respect to the regime of partial synchronization, the rates of energy exchange have diminished further, due to the much smaller intensity of the external forcing; cf.~Eqs.~(\ref{a1}) and (\ref{A2}). Otherwise, the overall behavior of $w_n^\Omega$, $w_n^\Gamma$ and $w_N^F$ is similar to that of partial synchronization. Namely, the rate of energy dissipation by friction (middle panel) exhibits a resonance peak for $K_n\approx 1$ and $\gamma_n \approx 0$, while $w_n^\Omega$ and $w_n^F$ show a change of sign near $K_n =1$, with a maximum and a minimum at each side. The origin of this resonance lies again in the dynamics of the individual deviations with respect to the $\Omega$ set, Eq.~(\ref{mea}). Indeed, the collective average motion of the ensemble, now added with oscillator $1$ and forced with unitary frequency from outside the system, acts as an excitation over each oscillator, inducing larger responses when the individual frequencies are closer to that of the external force. 

\section{Discussion and conclusion} \label{sec5}

We have studied the stationary collective dynamics of an ensemble of coupled mechanical phase oscillators subjected to an external harmonic excitation applied to one of the oscillators, and to friction forces. Heterogeneity in the system is given by different individual moments of inertia and friction coefficients. Our attention was focused on finding the rates at which energy is added from the external excitation and transferred between different parts of the ensemble, and the power dissipated by friction. To this end, we have proposed a series of approximations that allowed for analytical treatment of the equations of motion. Three regimes of collective motion were characterized, with different degrees of entrainment with the external excitation. Our approximations were validated by comparison of the rates of energy exchange derived from analytical results and from numerical integration of the equations of motion in the three regimes. 

When the external excitation is sufficiently strong, the ensemble becomes fully synchronized with the excitation and follows it rigidly. Even in this very simple form of coherent motion,  the rate of energy exchange between the oscillators not directly connected to the excitation varies its sign depending on how the individual moments of inertia and friction coefficients differ from their average values over the ensemble. 

Collective motion becomes richer when the entrainment with the external excitation is weaker, as we have shown in the regimes of partial synchronization and desynchronization. In these cases, the main dynamical feature is a resonance effect in which individual oscillators respond to the overall motion of the ensemble, which is in turn driven --but not completely entrained-- by the excitation. Superimposed to the uniform drift of its phase, each element performs an oscillation whose specific frequency is controlled by its moment of inertia.  Resonance occurs when this individual frequency approaches the frequency of the external harmonic forcing. In this situation, first, the power dissipated by friction increases abruptly when the friction coefficient decreases, as a consequence of the  growth in amplitude of individual oscillations. At the same time,  the individual rates of energy exchange between oscillators and with the external excitation switch their sign across the resonance. In other words, the flow of energy from each oscillator is positive or negative depending of its moment of inertia being above or below the average.

These results show, in a simplified scenario, how energy exchange between different parts of a system formed by coupled mechanical oscillators can be manipulated by tuning individual parameters, impacting on the overall dynamics of the ensemble. Although obtained within a theoretical context, such conclusions may be relevant to the design and operation of technological devices based, for instance, on fast damping for rapid switching between different functional regimes \cite{9,12} or where, on the contrary, sustained oscillatory motion at low dissipation rates is required \cite{10,11,low}.

\end{document}